\documentstyle[11pt,cs11,html,epsf]{article}

\begin{document}

\title{ Rotation and H$\alpha$~ Emission Above and Below the Substellar Boundary}

\author{G. Basri}
\affil{Astronomy Department, University of California,
    Berkeley, CA 94720}

\begin{abstract}

I present the results of a multiyear survey of very low mass stars and brown dwarfs, at high spectral resolution. The spectra were gathered with the HIRES echelle at the Keck Observatory. Some of these objects are stellar and others are substellar (or ambiguous). Early indications that such objects can be rapidly rotating but display little H$\alpha$~ emission turn out to be commonly true. This is the opposite of the relation between rotation and activity in solar-type stars. The H$\alpha$~ surface flux drops precipitously at the bottom of the main sequence, and seems to be related to the luminosity or temperature of the objects. There is a general trend to higher rotation velocities as one looks at objects of lower luminosity. I discuss several possible explanations for these results. The dynamos for these objects are probably fully turbulent, driven by convection, and thus more directly related to the object's luminosity. They may be quenched when the rotational velocities become too fast in comparison to the convective velocities (supersaturation). Another possibility is that the atmospheres of the cooler objects are becoming sufficiently neutral to decouple atmospheric motions from the field. Either of these could explain why young brown dwarfs can be magnetically active while older brown dwarfs are not. A final possibility is that instead of being quenched, the field configuration in rapid rotators changes to a less conducive form for dissipative heating. This could explain why flares are occasionally seen on generally inactive objects.

\end{abstract}


% 
% All my objects and aliases are after the Table
%

\section{Introduction}

One of the most studied properties of late-type stars is their magnetic activity. The quest to discover why stars of the same spectral type have such different levels of activity was answered in the 1980's with the ``rotation-activity'' connection. For solar-type stars, the faster the star spins, the greater is its total magnetic flux (usually measured through emission line proxies or X-ray luminosity). The explanation for this connection is via a magnetic dynamo, which relies on the Coriolis force for part of its operation. The Sun is thought to have such a dynamo operating at the base of its convection zone (a ``shell'' dynamo).
\index{Rotation-activity connection}

Even for the Sun, there is clearly another dynamo mode present (Title \& Schrijver 1998). The so-called ``distributed turbulent'' dynamo is thought to operate throughout the convection zone, and depend less on rotation. It is responsible for most of the flux present during solar minimum. As one moves to later spectral types, the convection zone deepens and the size of the shell  dynamo decreases. Thus one expects that by the time stars become fully convective, the turbulent dynamo will have taken over, and the disappearance of the shell dynamo is hardly noticeable. This occurs at about spectral type M3, and explains why activity levels do not suddenly change there (Giampapa et al 1996).
\index{turbulent dynamo}

The field emerges on smaller scales from the turbulent dynamo than the shell dynamo, and are less efficient at removing angular momentum. At about the fully convective boundary, the spindown time of stars begins to lengthen. Most early M field stars are found to be slowly rotating (except when young), and the incidence of measurable rotation picks up at about M3.5 (Delfosse et al 1998). The relation between rotation and activity still appears at mid-M in the sense that flare and dMe stars tend to be more rapid rotators. The fraction of stars which are dMe increases at later types; this is mostly because of the ``contrast effect'' whereby the same surface flux of a given hot diagnostic stands out better and better against the cooling photospheric continuum. Put another way, a constant emission line equivalent width represents less and less surface flux at cooler effective temperatures.
\index{dMe stars}

In the late M stars, the connection between rotation and activity is definitely changing. Basri and Marcy (1995) found an M9.5 star which has very rapid rotation, but no H$\alpha$~ emission. A second example was found by Basri et al (1996). Here we present the results of a multi-year survey of very late-type stars and brown dwarfs for rotation speeds and H$\alpha$~ emission. While most objects are stellar, some are confirmed brown dwarfs, and others may be substellar. The first part of this paper presents the results from that survey. In the second part we speculate a little about what the explanation for these results might be.

\section{Observational Results}

All observations were conducted with the HIRES echelle on the Keck I 10-m telescope. We used settings similar to those of Basri \& Marcy (1995). There were more than 50 targets, all M5 or later, including several examples of the new L spectral type. Some targets were in Basri et al (1996), and a few data have been added from the surveys of Basri et al (1996) and Tinney and Reid (1998). 
\index{Observatories!Keck}

\subsection{H$\alpha$~ Emission}

\begin{figure}
\plotone{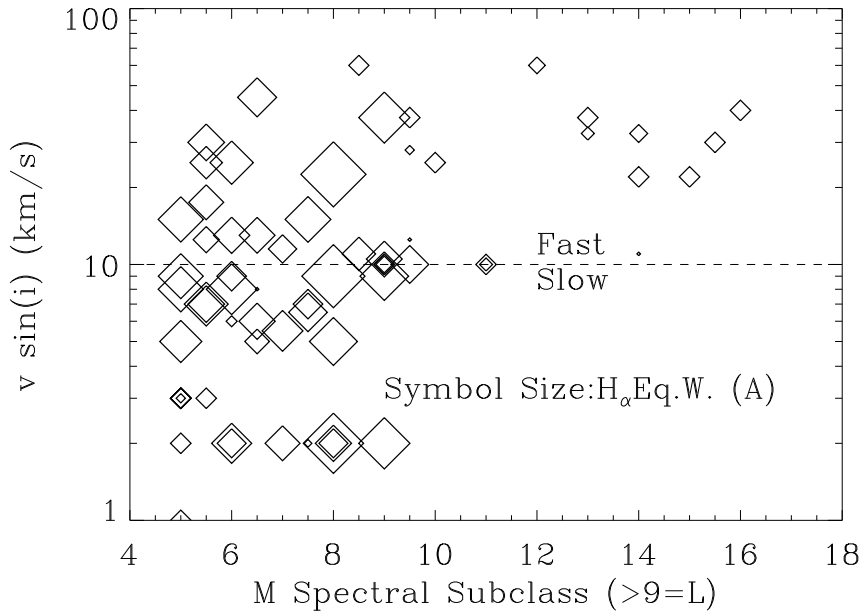} 
\caption{Rotation, Spectral Type, and H$\alpha$~ emission. Each ordinate shows a measured rotation velocity and the abscissa gives M spectral subclasses. For 10 and beyond these are L spectral subclasses (plus 10). The size of each point is scaled logarithmically with the H$\alpha$~ equivalent width (small points indicate low emission). The small symbols represent no detected emission at all.} \label{fig-1}
\end{figure}

 Because of the extremely cool photospheres for the stars in our sample, H$\alpha$~ can only show up if there is chromospheric or coronal heating. All our objects later than M5 and earlier than M9 show H$\alpha$~ emission. The fraction of stars with emission is know to be high after about M5 (Hawley, Gizis \& Reid 1996), although they find a peak fraction of 60\%. Some of the emission strengths we measure are near or below their observational limits. At about M9.5, we see a marked decrease in the emission, with several stars showing no emission and others with only weak lines. This data is listed in Table~\ref{tbl-1}, and  depicted in Fig.~\ref{fig-1}. There is a sudden drop in the fraction of active stars at about the end of the M sequence. The large equivalent widths are all in the late M stars (though not all stars are active), while the L stars are all inactive.
\index{M stars!H$\alpha$~ emission} 
\index{L stars!H$\alpha$~ emission}

A given value of emission equivalent width represents a dramatically weakening surface flux as we move into the late M and L stars.  The continuum, which defines the normalization of equivalent width, is dropping very quickly with temperature (H$\alpha$~ now occurs in the Wien part of the Planck function). We have used models by Allard and Hauschildt (see Mohanty et al, this volume) to estimate the conversion of equivalent width to surface flux. The models indicate that the continuum flux near H$\alpha$~ drops by almost a factor of 2 for each 100K in effective temperature.

\begin{figure}
\plotone{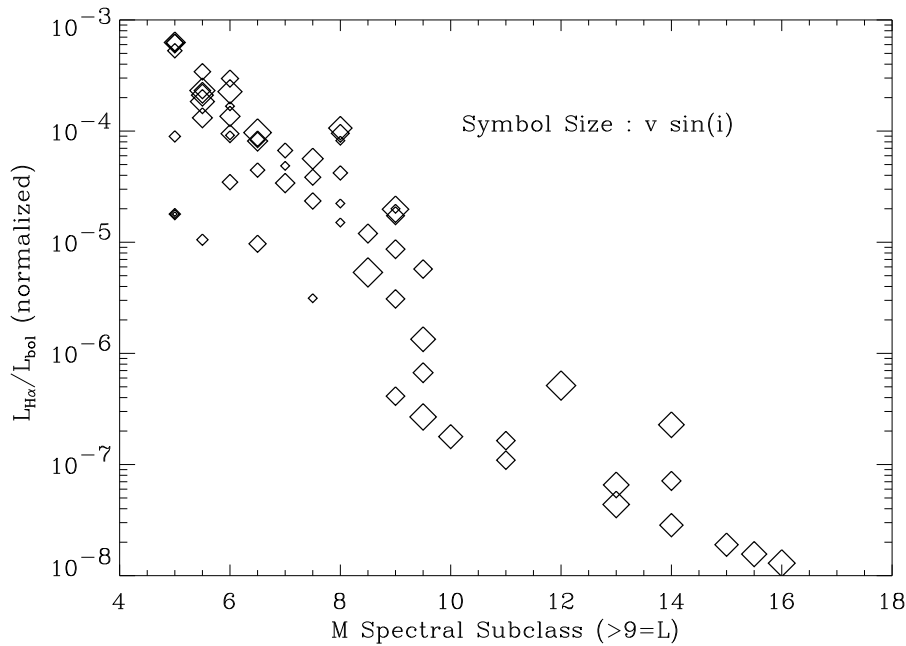} 
\caption{The relation between activity and spectral type. Unlike with solar-type stars, they are closely tied together. Cooler stars show much less activity (as measured either by luminosity ratio (as here) or by surface flux. The symbol sizes are scaled logarithmically by rotation velocity. The lower boundary of points are actually upper limits (emission not detected).} \label{fig-2}
\end{figure}

One of the main results of this paper is shown in Fig.~\ref{fig-2}. As was pointed out by Basri \& Marcy (1995), the fact that H$\alpha$~ equivalent widths remain fairly constant (and low) throughout this temperature range implies that the surface flux from heated plasma is dropping dramatically. It seems, in fact, that the drop in surface flux could be directly related to the temperature, or more likely but equivalently, the luminosity (since these dwarfs are all about the same size). To make it clear that the relative activity levels are falling, we show L$_{\rm{H}\alpha}/\rm{L}_{\rm{bol}}$ in Fig.~\ref{fig-2}. I have approximated L$_{\rm{bol}}$ by using our estimated temperature scale to spectral type conversion, and making a (small) correction for radius. With either surface flux or luminosity ratio, there is a dramatic decrease in the implied non-radiative heating (or amount of chromospheric plasma) as the objects get cooler and less luminous.

One might wonder whether there is some problem with H$\alpha$~ in cool dusty atmospheres that prevents it being in emission. The point is that we are considering chromospheric plasma, which by definition is hot enough to produce H$\alpha$~. The absence of emission thus implies that there is no such plasma present. There cannot be a substantial corona in the stars showing very weak H$\alpha$~ emission, because it would create a chromosphere by photoionization (Cram 1982) that would easily show up. I therefore conclude that the amount of stellar activity drops steeply at the bottom of the main sequence, and that its levels may be fairly directly tied to the stellar luminosity.

\subsection{Rotational Velocities}

We determine rotational velocities by cross-correlation with rotational standards, following the general procedures of Basri \& Marcy (1995). We used molecular features near 8500\AA for this purpose. For the M stars, GJ 406 was our standard. For the L stars we do not have any non-rotators in the sample, so we used the slowest one: 2MASS 1439+19. We checked its velocity through an intermediate calibration with LHS 2924. We carefully considered the effects of a calibrator with a finite velocity, and found that it does not substantially effect the determinations for $v$ sin($i$)~ 20 km/s or higher (slower cases may be overestimated, but are good to 5 km/s or better). Pressure broadening is not really a factor for these rapid rotators.
\index{*2MASS 1439+19}
\index{*GJ 406|LHS 36|CN Leo}

Down to about M3, there are rather few field stars with $v$ sin($i$)~ easily detectable (above 3 km/s). A good summary of the situation is given by Delfosse et al (1998). The situation is reversed in our sample: there are rather few stars too slow to be measurable. And this is despite the fact that $v$ sin($i$)~ is always a $lower$ limit on the true equatorial velocity. Our velocity determinations are listed in Table ~\ref{tbl-1}, and shown in Fig.~\ref{fig-1}. For convenience, we refer to stars with $v$ sin($i$)~ $<10$ km/s as ``slow'', and those with higher velocity as ``fast''. At M5 in our sample, there is only 1 star which is fast. For later subclasses, the fast stars are as numerous as the slow ones. It is very rare to find early field M stars with rotations higher than 20 km/s(and they are young and active), but there are several later than M5 and they are not particularly active. By the end of the M spectral class there are no slow stars left, and that remains true for all our L stars (indeed, we had trouble finding one slow enough to serve as a rotational calibrator).
\index{M stars!rotation velocity} 
\index{L stars!rotation velocity}

Is the rapid rotation intrinsic, or just a result of looking at young objects? There is a bias against finding old stars in the L class, as there is an increasing fraction of brown dwarfs there. These get even cooler and fainter with age, and so become harder to find. Those L stars that have been shown to be substellar by the lithium test have all been less than 1 Gyr old. Almost none of our M stars show lithium, so we can assume they are mostly stellar, and older. At the very least, it is clear that the average rotation velocity is much higher for the L stars than late M stars, and the late M stars are more rapidly rotating than the early M stars. Spindown times obviously increase as the mass of the objects decreases. For the L stars, it is striking that their velocities are so high, even before correcting for projection effects. It seems that the typical such object has a rotation period measured in hours; many of them are rotating faster than Jupiter.
\index{brown dwarfs!rotation velocity}

\begin{table}
\caption{Observed Rotation and H$\alpha$~ Emission.} \label{tbl-1}
\begin{center}\scriptsize
\begin{tabular}{ccrr}
Star & Spectral Type & $v$ sin($i$)~     & H$\alpha$~ Eq.W. \\
     &               &  (km/s) & (\AA) \\
\tableline
GJ 1093     &  M5.0&   3.0&   1.0 \\
GJ 1230B     &  M5.0&   $<$3.0&  $<$0.2 \\
GJ 1057     &  M5.0&   $<$3.0&  $<$0.2 \\
GJ 905     &  M5.0&   $<$3.0& -0.2 \\
GJ 810B     &  M5.0&   $<$3.0& -0.6 \\
GJ 3076     &  M5.0&   15. &   7.0 \\
GJ 3454      &  M5.0&   8.0&   7.0\\
GJ 1156     &  M5.0&   9.0&   6.9 \\
GJ 1154A     &  M5.0&   5.0&   5.9 \\
GJ 273      &  M5.0&   $<$3.0& -0.2 \\
GJ 65AB     &  M5.5 &   30. &   4.4 \\
LP 759-25    &  M5.5 &   12.5 &   2.5 \\
GJ 1245A     &  M5.5 &   17.5 &   4.0 \\
GJ 1245B     &  M5.5 &   7.0&   4.4 \\
GJ 1002      &  M5.5 &   $<$3.0&  $<$0.2 \\
GJ 1245A     &  M5.5 &   25. &   3.5 \\
YZ CMi     &  M5.5 &   7.0&   6.5 \\
GJ 3828B    &  M6.0&   6.0&   1.1 \\
GJ 412B     &  M6.0&   8.0&   9.4 \\
GJ 406      &  M6.0&   $<$3.0 &   5.3 \\
GJ 1111     &  M6.0&   13. &   4.3 \\
GJ 3622     &  M6.0&   $<$3.0&   2.9 \\
GJ 2005     &  M6.0&   9.0&   3.0 \\
CTI 2332+27    &  M6.0&   25. &   6.3 \\
LP 713-47    &  M6.5 &   13. &   4.2 \\
CTI 1747+28    &  M6.5 &   45. &   5.0\\
CTI 0156+28     &  M6.5 &   5.0&   2.3 \\
CTI 1539+28     &  M6.5 &   8.0&  0.5 \\
GJ 4281     &  M6.5 &   6.0&   4.3 \\
VB 8      &  M7.0&   5.5 &   5.5 \\
CTI 1156+28      &  M7.0&   11.5 &   2.8 \\
GJ 3877     &  M7.0&   $<$3.0&   4.0 \\
LHS 2645      &  M7.5 &   6.5 &   4.9  \\
LHS 2632     &  M7.5 &   $<$3.0&  0.4 \\
2MASS 1254+25    &  M7.5 &   15. &   7.2 \\
2MASS 1256+28    &  M7.5 &   7.0&   3.0\\
LP 412-31     &  M8.0&   9.0&   18.4 \\
RG 0050-2722     &  M8.0&   $<$5.0&   2.9 \\
LHS 2243      &  M8.0&   $<$5.0&   15.8 \\
VB 10      &  M8.0&   8.0&   4.3 \\
LHS 2397a    &  M8.0&   22.5 &   20.5 \\
2MASS 1242+29    &  M8.0&   5.0&   8.1 \\
TVLM 513-46546    &  M8.5 &   60. &   1.8 \\
CTI 1156+38     &  M8.5 &   11. &   3.7 \\
DENIS 2146-21    &  M9.0&   10. &  $<$0.2\\
LHS 2924     &  M9.0&   10. &   1.5 \\
CTI 0126+57     &  M9.0&   10.5 &   4.2 \\
TVLM 868-110639   &  M9.0&   37.5 &   9.6 \\
DENIS 1208+01    &  M9.0&   10. &   --\\
LHS 2065     &  M9.0&   9.0&   8.4 \\
BRI 1222-1222     &  M9.0&   2.0&   9.7 \\
ESO 207-61     &  M9.5 &   10. &   4.7 \\
DENIS 0021-42     &  M9.5 &   12.5 &  0.5 \\
LP 944-20     M9.5     &  M9.5 &   28. &   1.0\\
BRI 0021-0214    &  M9.5 &   37.5 &  $<$0.2 \\
DENIS 0909-06    &  L0. &   25. &  $<$0.2 \\
G196-3B      &  L1. &   10. &  $<$0.2 \\
2MASS 1439+19   &  L1. &   10. &  $<$0.3 \\
Kelu-1     &  L2. &   60. &   1.5 \\
GD 165B      &  L3. &   37.5 &  $<$0.2 \\
DENIS 1058-15    &  L3. &   32.5 &  $<$0.3 \\
2MASS 1146+22   &  L4. &   22. &  $<$0.2 \\
DENIS 1228-15    &  L4. &   11. &  0.5 \\
LHS 102B     &  L4. &   32.5 &   1.6 \\
DENIS 0205-11    &  L5. &   22. &  $<$0.2 \\
2MASS 1632+19   &  L5.5 &   30. &  $<$0.2 \\
DENIS 0255-47    &  L6. &   40. &  $<$0.2 \\
\end{tabular}
\end{center}

%\tablenotetext{a}{Sample footnote for Table~\ref{tbl-1}}
\end{table}

\index{*GJ 1093|LHS 223}
\index{*GJ 1230B|LHS 3404}
\index{*GJ 1057|LHS168}
\index{*GJ 905|LHS 549}
\index{*GJ 810B|LHS 500}
\index{*GJ 3076|LP 467-16}
\index{*GJ 905}
\index{*GJ 3454}
\index{*GJ 1156}
\index{*GJ 1154A}
\index{*GJ 273}
\index{*GJ 65AB}
\index{*LP 759-25}
\index{*GJ 1245B}
\index{*GJ 1245A}
\index{*GJ 1002}
\index{*GJ 1245A|LHS 3494}
\index{*YZ CMi|GJ 285|LHS 1943}
\index{*GJ 3828B|LHS 2876}
\index{*GJ 412B}
\index{*GJ 406|LHS 36|CN Leo}
\index{*GJ 1111|LHS 248}
\index{*GJ 3622|LHS 292}
\index{*GJ 2005|LHS 1070}
\index{*CTI 2332+27}
\index{*LP 713-47}
\index{*CTI 1747+28}
\index{*CTI 0156+28}
\index{*CTI 1539+28}
\index{*GJ 4281|LHS 523}
\index{*VB 8|GJ 644C|LHS 429}
\index{*CTI 1156+28}
\index{*GJ 3877|LHS 3003}
\index{*LHS 2645|LP 218-8}
\index{*LHS 2632|LP 321-222}
\index{*2MASS 1254+25}
\index{*2MASS 1256+28}
\index{*LP 412-31|LP 315-53}
\index{*RG 0050-2722}
\index{*LHS 2243}
\index{*VB 10|GJ 752B|LHS 474}
\index{*LHS 2397a}
\index{*2MASS 1242+29}
\index{*TVLM 513-46546}
\index{*CTI 1156+38}
\index{*DENIS 2146-21}
\index{*LHS 2924}
\index{*CTI 0126+57}
\index{*TVLM 868-110639}
\index{*DENIS 1208+01}
\index{*LHS 2065}
\index{*BRI 1222-1222}
\index{*ESO 207-61}
\index{*DENIS 0021-42}
\index{*LP 944-20}
\index{*BRI 0021-0214}
\index{*DENIS 0909-06}
\index{*G196-3B}
\index{*2MASS 1439+19}
\index{*Kelu-1}
\index{*GD 165B}
\index{*DENIS 1058-15}
\index{*2MASS 1146+22}
\index{*DENIS 1228-15}
\index{*LHS 102B}
\index{*DENIS 0205-11}
\index{*2MASS 1632+19}
\index{*DENIS 0255-47}

\subsection{The Rotation-Activity Connection}

For solar-type stars, a clear increase in magnetic activity is seen with an increase in rotation velocity. A nice summary figure for the rotation-activity connection can be found in Randich 1998 (see also Saar, this volume). The variables in which this relation is best expressed are either surface flux vs. rotation period, or luminosity ratio (with L$_{bol}$) vs. Rossby number. The Rossby number is itself the ratio of convective overturn time to rotation period. Fig.~\ref{fig-1} shows that the activity levels do not increase with rotation velocity. There seems to be little connection between the two at low velocities, and none of the rapid rotators have high emission levels. This is without regard to the spectral type.

The dependence on spectral type can be seen in Figs.~\ref{fig-1} and \ref{fig-2}. They show that the L stars are both rapidly rotating and inactive, while late M stars span a range of rotations but are more active. The L star sample is substantially smaller, so it is possible that more slowly rotating or more active examples will be uncovered. But it is already clear that they are different. Based on the lithium test, at least a third of them are brown dwarfs. This fraction increases toward later L types, and becomes 100\% by about L4 or L5 (where we think the minimum main sequence temperature lies). 
\index{Rotation-activity connection}

The general conclusion that can be drawn from this is that the rotation-activity connection is weakened once one crosses the fully convective boundary, and that it completely disappears (and perhaps even reverses) cooler than M9. Since spindown is due to angular momentum loss via magnetic winds, it is not surprising that low activity levels go along with rapid rotation. To the extent that some of our objects are on the main sequence, they are probably old and show that there is little spindown over long times. In order to test older brown dwarfs, we will have to measure rotation velocities for the methane dwarfs. 
\index{magnetic braking}

\section{Possible Explanations for the Results}

There are several possible explanations for these results. The truth may involve more than one of them (or another not discussed here). One is that the ionization levels in the photosphere may have become so low that there is insufficient conductivity to allow coupling of the magnetic field to the gas.  Then gas motions may not twist up or excite waves in the fields, and there is no dissipation to heat the upper atmosphere.  This has to be true even in the face of ambipolar diffusion, which couples small numbers of ions to the neutrals fairly effectively (as in T Tauri disks).  The alkali metals, which are the last source of electrons, are becoming quite neutral in the L stars.  A possible counter-example to this hypothesis is provided by the detection of (non-flaring) H$\alpha$~ emission in a methane dwarf, which should be $very$ neutral (Liebert et al, this volume). Another potential wrinkle is that the plasma is hotter beneath the photosphere and will become ionized at some depth. There will be convective motions there too, so it is not obvious how the surface fields will behave.

The traditional rotation-activity connection may arise because activity increases with decreasing Rossby number. In solar-type stars, activity levels increases steadily from a Rossby number of 5 down to 0.1.  Coronal activity ``saturates'' for Rossby numbers between about 0.1-0.01; the activity flattens out at its maximal values seen. It has been shown that most activity diagnostics scale simply like L$_x$/ L$_{bol}$ (Schrijver 1987). 
All low mass objects should have turbulent dynamos, which are driven by convective motions.  Rotation can enhance production of fields, and the amplitude of convective velocities also does.  But convective overturn times scale with effective temperature in these objects.  At the bottom of the main sequence they can increase to months, while typical spin periods are dropping to hours. A simplistic analysis using mixing length theory suggests that the Rossby numbers for stars in our sample lie mostly in the range 0.01-0.001. This is because the convective overturn times scale down with effective temperature, and the rotation velocities get higher for the cooler objects.   
\index{Rossby number}

As first noted by Vilhu \& Rucinski (1983) and nicely summarized by Randich 1998, there is some evidence for ``supersaturation'' - an actual turndown of activity levels -- for Rossby numbers less than 0.01. I speculate that the dynamo may be unable to operate efficiently at such small Rossby numbers, perhaps because rotation organizes the flows too much. A possible counterexample to this hypothesis is provided by the very rapid rotator Kelu-1, which exhibits a persistent (though rather weak) H$\alpha$~ emission line. This hypothesis can be tested when a sufficient sample of L stars with H$\alpha$~ emission has been collected. They should show systematically lower rotational velocities if rotation is important in this way. 
\index{supersaturation}
\index{*Kelu-1}

The only population of brown dwarfs that is known to be fairly active is the very young ones. Those found in star forming regions (eg. Wilking et al 1999) or as old as the Pleiades (eg. Zapatero-Osorio et al 1997) tend to show reasonably strong H$\alpha$~, or even X-ray emission (Neuhauser et al 1998). These are also of course more luminous, and perhaps their convective velocities are still high enough to prevent supersaturation. The Pleiades objects have been shown to be generally rapid rotators (Oppenheimer et al 1998).
\index{brown dwarfs!magnetic activity}

A related possibility is that the field is not actually quenched by rapid rotation, but instead takes on a relatively stable, low multipole character like that of Jupiter.  In that case, the field might be sufficiently quiet (especially in conjunction with low conductivity) that it does not suffer the dissipative configurations that power stellar activity. Thus, the objects might still have strong fields, but little stellar activity.  This could be tested in principle using Zeeman diagnostics. Valenti (this volume) suggests using FeH for objects in this temperature range, and have shown that it can work in late M stars. A suggestion that there are indeed fields comes from the fact that flares have been seen in objects that seem otherwise quiescent, such as VB10 (Linsky et al 1995) and 2MASSW J1145572+231730 (Leibert et al 1999).  
\index{*2MASSW J1145572+23173}
\index{*VB 10|GJ 752B|LHS 474}

It is clear that a new regime in the study of stellar activity has been reached at the bottom of the main sequence. This will be the motivation for much additional work in the coming years as our sample of very cool objects is greatly increased by the 2MASS, DENIS, and Sloan sky surveys. We will have to follow up with low and high dispersion spectroscopy and synoptic photometry to gather the data needed to fully sort this problem out. But in the process, we are likely to gain valuable new information on both stellar dynamos and the process of non-radiative heating that produces stellar activity.

\acknowledgments

I acknowledge the support of NSF grant AST96-18439 in this research. The data discussed was mostly gathered at the Keck Observatory, which is operated jointly by UC and Caltech. I am indebted to G. Chabrier for suggesting the possibility of a changed field configuration due to rapid rotation. Subu Mohanty helped in the determination of some rotational velocities. 

\begin{question}{C. Bailer-Jones}
I am a bit concerned about your $v$ sin($i$)~ measurements. How do you deduce $v$ sin($i$)~ in the absence of a well-known slow enough and late enough standard?
\end{question}
\begin{answer}{G. Basri}
(This is commented on in the manuscript) We worried about this quite a bit. Originally we developed a means of correcting the answer using a moderately fast rotator. But then we found one at 10 km/s which is plenty slow enough for most of the L stars.
\end{answer}
\begin{question}{R. Jefferies}
Are the phenomena you have presented related to the question of the period gap in CVs?
\end{question}
\begin{answer}{G. Basri}
Yes, they probably are. The same sort of conductivity arguments have been invoked recently by the Meyers for the period gap systems, and the low mass secondaries are presumably similar to the objects I discussed. Indeed, the rotation periods are sometimes near 2 hours as well.
\end{answer}

\end{document}